\documentclass{ptapap}

\author{E. Niemczura}[niemczura@astro.uni.wroc.pl, UWR]
\author{A. Pigulski}[UWR]
\author{H. Lehmann}[TLS]
\author{K. Kami\'{n}ski}[UAM]
\author{G. Catanzaro}[OAC]
\author{I. Stateva}[ISB]
\author{M. Napetova}[ISB]
\author{BEST}[BEST]

\affil[UWR]{Astronomical Institute, University of Wroc\l{}aw, Kopernika 11, 51-622 Wroc\l{}aw, Poland}
\affil[TLS]{Th\"{u}ringer Landessternwarte Tautenburg, Sternwarte 5, D-07778 Tautenburg, Germany}
\affil[UAM]{Astronomical Observatory Institute, Faculty of Physics, A. Mickiewicz University, S\l{}oneczna 36, 60-286 Pozna\'{n}, Poland}
\affil[OAC]{INAF - Osservatorio Astrofisico di Catania, Via S.Sofia 78, I-95123 Catania, Italy}
\affil[ISB]{Institute of Astronomy with NAO, Bulgarian Academy of Sciences, Sofia 1784, Bulgaria}
\affil[BEST]{Bright Target Explorer (BRITE) Executive Science Team}

\title{Photometric and spectroscopic variability of 53 Per}

\begin{document}

\maketitle

\begin{abstract}

A new investigation of the variability of the SPB-type star 53\,Per is presented. 
The analysis of the BRITE photometry allowed us to determine eight independent frequencies and the combination one.
Five of these frequencies and the combination one were not known before.
In addition, we gathered more than 1800 new moderate and high-resolution spectra of 53\,Per spread over approximately six months. 
Their frequency analysis revealed four independent frequencies and the combination one, all consistent with the BRITE results.

\end{abstract}


53\,Per (HD\,27396, B4\,IV, V\,=\,4.85\,mag) is one of the first studied variable mid-B stars of the northern hemisphere. 
Its photometric and spectroscopic data have been gathered and investigated from the mid seventies 
\citep[see][and references therein]{1998A&A...331.1046C, 1999A&A...341..574D}. 
Nonetheless, our understanding of the pulsational behaviour of 53\,Per is still far from complete.
The star was a prototype of a group of mid-B type stars showing variability due to non-radial pulsations in line profiles, 
later incorporated into SPB class of variable. Before the BRITE observations three modes were known in this star \citep{1999A&A...341..574D}.\\

\vspace{0.5cm}

53\,Per was one of the stars observed by BRITE satellites in the Perseus field. 
The observations spanned $170$ days between September $2$, $2014$, and February $18$, $2015$, and come from a single BRITE satellite, UniBRITE (UBr). 
Details of the instrumentation and observing procedure are given by \citet{2014PASP..126..573W}.
The combined red-filter light curve consists of $165\,274$ data points.
Prior to time-series analysis BRITE photometry was corrected for instrumental effects. 
The corrections included outlier rejection and decorrelations with centroid positions and CCD temperature. 
Final correction accounted for offsets between four individual setups of the UBr observations.
The obtained light curve was used for time-series analysis, which consisted of the calculation of the Fourier frequency spectrum,
identification of the highest maximum in the spectrum, and pre-whitening the original light curve with all previously detected frequencies. 
In total, $8$ significant independent frequencies were identified in BRITE photometry.
The highest amplitudes were obtained for the three previously identified modes ($f_1=0.46112$, $f_2=0.59388$, $f_3=0.47116$\,d$^{-1}$), with 
amplitudes ranging from about $7$ to over $20$\,mmag. For the other modes, the amplitudes are lower than $5$\,mmag.
In addition, one combination frequency, $f_1+f_2$ was found. \\

\vspace{0.5cm}

The spectroscopic observations of 53\,Per lasted from October $2015$ to March $2016$. The moderate and high-resolution spectra 
were obtained with four different \'{e}chelle spectrographs, working at the $2$-m telescope of the Th\"{u}ringer Landessternwarte Tautenburg (TLS, $1428$ spectra), 
the $0.5$-m Pozna\'{n} Spectroscopic Telescope (PST1, 322), the $91$-cm telescope of the Osservatorio Astrofisico di Catania (OAC, $16$), 
and the $2$-m telescope of the Bulgarian National Astronomical Observatory (BNAO, $24$).
The spectra have an average S/N ratio of about $100$. After the standard reduction and calibration procedures all spectra were normalised to the local continuum.
To compute radial velocities (RV), we extracted useful neutral helium lines ($4713.15$, $4921.93$, $5875.62$, and $6678.16$ \AA{}).
RVs were calculated with using the cross-correlation method. 
In total, $4$ significant independent frequencies and the combination one ($f_1+f_2$), were identified.
Two frequencies with the highest amplitudes were known before ($f_1$ and $f_2$). 
All spectroscopically obtained frequencies are consistent with the BRITE photometry analysis.\\

\vspace{0.5cm}

The investigation of BRITE photometry has enabled us to identify more frequencies than was known before, including the combination one.
Thanks to this, the next step of our analysis will be mode identification and seismic analysis of the star  
(Niemczura et al., in preparation).

\acknowledgements{The calculations have  been  carried  out  in  Wroc\l{}aw  Centre  for  Networking and  Supercomputing  (http://www.wcss.pl),  
grant No. 214. EN is supported by NCN through research grant No. 2014/13/B/ST9/00902.
HL is supported by DFG grant LE 1102/3-1.
KK acknowledges NCN research grant No. 2011/01/D/ST9/00427.
Based on data collected by the BRITE Constellation satellite mission, designed, built, launched, operated and supported by the Austrian 
Research Promotion Agency (FFG), the University of Vienna, the Technical University of Graz, the Canadian Space Agency (CSA), the University of 
Toronto Institute for Aerospace Studies (UTIAS), the Foundation for Polish Science \& Technology (FNiTP MNiSW), 
and National Science Centre (NCN).
}

\bibliographystyle{ptapap}
\bibliography{ENiemczura-Innsbruck2016.bib}

\end{document}